\documentclass[pre,amsmath,amssymb,twocolumn,superscriptaddress]{revtex4}

\usepackage{amsmath,amssymb}
\usepackage[usenames]{color}
\usepackage{amssymb}
\usepackage{dsfont}
\usepackage{grffile}
\usepackage[pdftex]{graphicx}
\usepackage{amsmath, amstext, amssymb, amsfonts, amsxtra}
\usepackage{textcomp}
\usepackage{xspace}
\DeclareSymbolFontAlphabet{\amsmathbb}{AMSb}

\newcommand{\cur}{\mathcal{J}}

\newcommand{\ave}[1]{\langle #1 \rangle }

\newcommand{\rhop}{{\rho}}

\newcommand{\Hop}{{H}_{S}}
\newcommand{\sop}{{\sigma}}
\newcommand{\Dop}{\mathcal{D}}

\newcommand{\tr}{{\rm tr}}

\newcommand{\im}{{\rm i}}

\newcommand{\An}{{A}_{L}(\omega)}
\newcommand{\Ad}{{A}_{L}^{\dagger}(\omega)}

\newcommand{\w}{\omega}

\newcommand{\kbc}{k_B}

\begin{document}

\title{Perfect Thermal Rectification in a  Many-Body Quantum {{Ising}} Model}

\author{Emmanuel Pereira}
\affiliation{Departamento de F\'{\i}sica--Instituto de Ci\^{e}ncias Exatas, Universidade Federal de Minas Gerais, CP 702, 30.161-970 Belo Horizonte MG, Brazil}

\begin{abstract}
We address a keystone problem for the progress of phononics: the proposal of efficient thermal diodes. Aiming the disclosure of an easy itinerary for the building of a heat rectifier, we investigate unsophisticated
systems linked to simple thermal baths, precisely, {{ asymmetric quantum Ising models, i.e., simple quadratic models, involving only one spin component}}. We analytically show the occurrence of thermal rectification for the case of a chain with interactions long enough to connect the first to the last site. Moreover,
we describe cases of a perfect rectification, i.e., finite heat flow in one direction, and zero current in the opposite direction. We argue to indicate that the ingredients for the rectification are just given by  the quantum nature of the baths and dynamics, and by the structural asymmetry of the system, here in the inter-site interactions.
We believe that the description of a perfect thermal rectification in a simple many-body quantum model, that is, the presentation of a simple itinerary for the building of a diode shall stimulate  theoretical and experimental research on the theme.

\end{abstract}


\maketitle

\section{Introduction}

 Conduction by electricity and by heat are two key mechanisms of energy transport, but with different status in science.  In the one hand, modern electronics exhibits an amazing progress, with impact in our daily lives.
But on the other hand, phononics, the counterpart of electronics devoted to the control and manipulation of the heat current, is advancing in a walking pace. The reason for such a marasmus is the absence of a reliable and efficient thermal diode,
in contrast with the successful  electrical diodes and based nonlinear solid-state devices, such as electrical transistors.

In this context, aiming the proposal of suitable thermal diodes, in the present work we turn to the ``starting point'' of such an investigation and focus on the minimal ingredients necessary for the occurrence of
 thermal rectification. Besides the necessary ingredients, we also search for the possibility of a huge thermal rectification.

In most of the cases, at least in studies without the presence of special or elaborate baths, structural asymmetry, of course, and anharmonicity in the interaction (i.e., nonlinearity in the dynamics) have been considered the indispensable elements for the occurrence of thermal rectification
\cite{comentarios, SegalNitzan, Casati1, Casati2, WuSegal, LiRMP}.
In fact, the first proposals of thermal diodes \cite{Terraneo, LiC} were given by the sequential coupling of different segments with anharmonic (nonlinear) terms. However, they present serious problems, such as a small rectification factor that rapidly decays to zero as the system size increases.

The requirement of anharmonicity is true, at least for the most familiar models  for the study of heat conduction in insulating solids, namely, for classical chains of
oscillators, recurrently used since Debye \cite{Debye} and Peierls \cite{Peierls}. There is no rectification in any asymmetric version of a chain of classical harmonic oscillators. The same follows for the quantum chain of
oscillators. {{A transparent proof is given, e.g., by
the Landauer formula for the heat conduction as derived in Ref.\cite{DD}: the formula is symmetric under the interchange of the leads. See also Ref.\cite{OJ} for a more general Landauer formula and related properties.}}
Even for the self-consistent extension, i.e., for harmonic classical
systems with inner stochastic baths, the absence of thermal rectification is rigorously proved  \cite{PLA}. In this extended model,
the inner stochastic reservoirs describe only some mechanism of phonon scattering,  a ``residual'' effect of the anharmonicity absent in the potentials, but  represented by noise in the system. This model obeys the Fourier's law of heat conduction \cite{BLL, PF}, in contrast with the purely harmonic chain of oscillators, which means that the inner noise keeps, indeed, some footprint of anharmonicity. Interestingly, in the
quantum version of the self-consistent harmonic chain, that also obeys Fourier's law  \cite{DR}, rectification holds \cite{P1, BD}.

{{Considering the context of quantum spin systems, in Ref.\cite{Landi}, for the homogeneous $XXZ$ spin $1/2$ model and target polarization at the boundaries, the authors prove the absence of spin current rectification in the case of zero asymmetry parameter $\Delta$ (coefficient of $\sigma_{i}^{z}\sigma_{i+1}^{z}$), even in the presence of an
asymmetric external magnetic field. The spin rectification is present only for $\Delta \neq 0$. Recalling that the $XXZ$ model can be mapped into a problem of hard-core bosons involving
 creation and annihilation operators with quadratic potentials and an extra quartic term which is
proportional to $\Delta$ (Tonks-Girardeau model), the absence of rectification due to the vanishment of the quartic term is compared with the problem of classical oscillators, where rectification does not hold in the absence of
anharmonicity (i.e., in the absence of the quartic or higher order terms  in the potential of the classical problem of oscillators). The quantum Ising model involves only one spin component. In this context, one might say, in an abusive nomination, that the quantum Ising model,  to be treated here, seems to be a kind of harmonic part of the $XXZ$ model.}}

We need to remark that within specific approaches such that in a scheme involving  proper chosen baths, precisely, baths given by spin reservoirs with properly chosen magnetic fields, rectification has been described in some simple quadratic spin chains \cite{Arrachea}.

There are many other attempts to build thermal diodes by involving intricate schemes or complicate anharmonic interactions, for example, carbon nano-structures with elaborate shapes and asymmetries \cite{nano1}, including graphene nanoribons
\cite{Hu, nano2}. In particular, the first experimental work aiming the construction of a thermal rectifier was in a asymmetrically mass-load carbon nanotube \cite{Chang}, an interesting device, however, with a small rectification factor.

Our strategy here is the opposite, i.e., we want to get rid of intricate schemes and elaborate mechanisms. We turn to the analytical study of simple models in the search of the ingredients for heat rectification.
 We start from a quite simple, naked quantum model, namely, the quantum Ising model, pacifically coupled to thermal baths, without any intricate, special coupling, interaction or bath. Then we sequentially introduce small modifications, such as asymmetries and quadratic interactions beyond
nearest-neighbors, increasing the complexity, trying to provoke the onset of thermal rectification to find
the minimal ingredients. After that, we investigate the possibility of a huge rectification. We show that there is rectification in these simple many-body, asymmetric quantum Ising models if the interactions are long enough to couple the first to the last
site. Moreover, for an adequate choice of the parameters, we show the occurrence of a perfect rectification in the limit of zero temperature in one side of the chain: the heat current, which exists in one direction, vanishes as we invert the thermal baths, that is, when we try to invert the flow direction.

We offer an explanation for the rectification phenomenon. In the used model, we see that the quantum nature of the baths brings a temperature dependence into the bulk of the system. This effect together with a structural asymmetry in the chain (here, in the interparticle interaction) lead to rectification.
Precisely, when we invert the chain between two baths, the structure of the chain changes, and so the temperature distribution in the bulk, which depends on the fixed baths and also on the structure of the chain. Consequently,
the expressions for the heat current change and
rectification appears. No  intricate potential or specific heat bath is necessary for the phenomenon occurrence. We understand that the long range interaction is only necessary to avoid the vanishment of the heat current (possibly due to some hidden symmetry in this specific model). It shall not be, we believe,  a necessary condition for rectification in other related models.

A further comment. It is worth to stress that our results, involving transport in open quantum systems, interest to phononics and many other communities: nonequilibrium statistical physics, condensed matter, ultracold atoms, etc.

\section{Model}

 It is important to investigate genuine quantum models for many reasons: the present ambient of miniaturization due to the advance of nanotechnology and lithography, specific effects in low temperatures, etc. Here,
 we are, in some way,
stimulated by recent findings involving interesting rectification properties of asymmetric quantum spin $XXZ$ chains \cite{Prap17, PerfDiode}, the archetypal model of open quantum systems \cite{BP}.
Precisely, in some previous works involving boundary driven $XXZ$ models, with target polarization at the
edges, we found cases of an unique way direction for the energy current \cite{Prap17}, and a case of a perfect spin diode \cite{PerfDiode}. These findings make auspicious the investigation of heat rectification in the
$XXZ$ chain linked to real thermal baths. However, this model is  very intricate, with many effects and different properties according to the variation of its inner parameters. And so, its investigation seems to be completely
inadequate for the identification of the minimal ingredients. Then, trying to preserve the $XXZ$ rectifying property, we take, say, a simplified component of this Heisenberg family, the quantum Ising model. Precisely, we start from the $1/2$ quantum spin system, with Hamiltonian of the type
\begin{equation}
H_{S} = \sum_{i=1}^{N} h_{i}\sigma_{i}^{z} + \sum_{i,k} \Delta_{i,k} \sigma_{i}^{z}\sigma_{k}^{z}~,
\end{equation}
where $\sigma_{i}^{z}$ is the $z$ Pauli matrix at site $i$;
more specifications are described ahead, see Eqs.(\ref{H1}, \ref{H2}, \ref{H3}, \ref{H4}). The system is coupled to different baths at right (R) and left (L) sides. We assume the spin-boson coupling model in the $x$ component, i.e., we take the following Hamintonians for baths and system-baths interaction
\begin{eqnarray*}
H_{{\rm syst.-bath}}^{(n)} &=& \sigma^{x}_{n}\sum_{k} \xi_{k}\left( a_{k}^{(n)} + a_{k}^{(n) \dagger}\right)~,\\
H_{\rm bath}^{(n)} &=& \sum_{k} \omega_{k}a_{k}^{(n) \dagger}a_{k}^{(n)}~,
\end{eqnarray*}
where $n$ is $L$ or $R$; $\xi_{k}$ is the coupling strength of both baths; $a_{k}$ and $a_{k}^{\dagger}$ are the annihilation and creation operators of the boson mode $k$.
Such a modeling implies that the baths can flip only an unique spin at a time: always the first or the last spin of the chain system, i.e., the spin of the site linked to the bath. Transitions
simultaneously flipping  more than one spin (both spins, for example) are forbidden. We perform the microscopic derivation to arrive at the Lindbladians for the dynamics given in terms of the eigenfunctions of the
full system Hamiltonian $H_{S}$. In the Born-Markov approximation (applied to the baths manipulation), see \cite{BP}, the Lindblad master equation becomes, for $\hbar = 1$,
\begin{align}\label{mastereq}
  \frac{d\rho}{d t}=-\im[H_{S},\rho]+\Dop_L(\rho)+\Dop_R(\rho).
\end{align}
Here,  $\rho$ is the density matrix of the system, and $\Dop_L$ (similarly for $\Dop_R$) is the dissipator due to the coupling of the left (right) site with the left (right) bath, given by
\begin{align}
\Dop_L(\rhop)=&\sum_{\w>0}G(\w) \left\{ \left[1+n_L(\w)\right] \left[\An\rhop\Ad \right.\right. \nonumber \\
& \left.- \frac{1}{2} \left(\Ad\An\rhop + \rhop\Ad\An  \right)  \right]    \nonumber \\
&+n_L(\w) \left[\Ad\rhop\An \right.   \nonumber \\
& \left. \left. - \frac{1}{2} \left(\An\Ad\rhop + \rhop\An\Ad\right)  \right]\right\}~,
\end{align}
 where $\w=\epsilon_{k}-\epsilon_{i}$ is the energy difference between the two eigenstates $|\epsilon_i\rangle$ and $|\epsilon_k\rangle$ of $H$; $n_L(\w)=[\exp(\hbar\w/\kbc T_{L})- 1]^{-1}$ is the Bose-Einstein distribution for the heat bath, and $\kbc$ is the Boltzmann constant, which is taken as $1$ in what follows (such as $\hbar$). The Lindblad operator  $A_{L}(\w) = \sum_{\w} |\epsilon_i \rangle|\langle \epsilon_i |\sigma^{x}_{L}|\epsilon_k\rangle\langle \epsilon_k|$ gives the transitions induced by the bath. We assume an Ohmic bath, and so, $G(\w)=\lambda \w$, for both $L$ and $R$ reservoirs.

 We need to stress that such a simple quantum Ising model is an old and recurrently studied system. In particular, we recall that the case of $N=2$, i.e., the case of a junction (not a ``many-body'' model), is detailed investigated in Ref.\cite{Werlang}, with the same baths considered here. There, the authors also show a perfect rectification for their junction model. Here, however, we will show that the direct extension of this Ising model to $N>2$ leads to a system with no heat flow in the steady state, i.e., if we take an Ising  Hamiltonian with nearest neighbor interactions, then the heat current is zero in the steady state for $N>2$. Sometimes, for spin models and even other systems, there are drastic differences between $N=2$
 and $N>2$. For example, in Ref.\cite{Landi}, in the $XXZ$ model with target polarization at the boundaries and with  an asymmetric external magnetic field, the authors show the occurrence of spin rectification for $N>2$, but
 such a rectification does not appear for $N=2$.
 Here, we also have a difference between $N=2$ and other values, but, in some way, an opposite situation: heat current for $N=2$ and no current for $N>2$. But we will show that a modified and more complicated model (still quadratic and involving only one spin component) with long range interactions presents a nonvanishing heat current and also a perfect rectification.

\section{Results}

   To obtain the expression for the heat current $\cur$, we turn to the continuity equation,
\begin{align}
\frac{d\ave{H_{S}}}{dt}= -\nabla\cur = \cur_L - \cur_R.
\end{align}
From the master equation (\ref{mastereq}), we have
 \begin{align}
\frac{d\ave{H_{S}}}{dt}=\mathrm{tr}\big[\Hop\Dop_L(\rhop)\big]+\mathrm{tr}\big[\Hop\Dop_R(\rhop)\big].
\end{align}
In the steady state, ${d\ave{\Hop}}/{dt}=0$, and the heat current becomes $\cur= \cur_{L}= \cur_R$,
 \begin{align}
\cur_{L}=\mathrm{tr}\big[\Hop\Dop_L(\rhop)\big];~~~~\cur_{R} =-\mathrm{tr}\big[\Hop\Dop_N(\rhop)\big].
\end{align}
Moreover, the density matrix is diagonal in the energy eigenstates, and the Lindblad master equation is reduced to
\begin{equation}\label{LMEs}
\frac{d\rho_{jj}}{d t}= 0 = \Dop_L(\rho)_{jj} + \Dop_R(\rho)_{jj}.
\end{equation}

To follow with the computation, we completely specify our model. We first take the ``smallest many body'' system, a chain with $N=3$, i.e., a chain with one spin coupled to the left bath, another one at the different edge coupled to the right reservoir, and with the third one describing the ``bulk'' of the system. As said before, the simplest case $N=2$, which is a junction, was already studied in Ref.\cite{Werlang}. Interestingly, such a simple junction model is the basic
component of the quantum thermal transistor built in Ref.\cite{Miranda}. However, we repeat,
 sometimes there are drastic differences between $N=2$ and $N>2$, see, e.g., Ref.\cite{Landi}.

The Hamiltonian of our first case is
\begin{align}
H_{S} = h(\sop^{z}_{1} + \sop^{z}_{2} + \sop^{z}_{3}) + (\Delta + \delta)\sop^{z}_{1}\sop^{z}_{2} + (\Delta - \delta)\sop^{z}_{2}\sop^{z}_{3}~, \label{H1}
\end{align}
i.e., we take a system with uniform magnetic field and asymmetric interaction. It is immediate the computation of the eigenstates and eigenvalues. For a first analysis, let us take $h < \delta < \Delta$. In crescent energy  order, using $+$ and $-$ for the spin eigenvalues, we have
$|1\rangle = |-+-\rangle$, $E_{1} = -h-2\Delta$; $|2\rangle = |+-+\rangle$, $E_{2} = +h-2\Delta$; $|3\rangle = |+--\rangle$, $E_{3} = -h-2\delta$; $|4\rangle = |-++\rangle$, $E_{4} = +h-2\delta$;
$|5\rangle = |--+\rangle$, $E_{5} = -h+2\delta$; $|6\rangle = |++-\rangle$, $E_{6} = +h+2\delta$; $|7\rangle = |---\rangle$, $E_{7} = -3h+2\Delta$; $|8\rangle = |+++\rangle$, $E_{8} = +3h+2\Delta$.

Then, we turn to the steady state, i.e., to Eq.(\ref{LMEs}). We precisely write these equations  in terms of the transition rate of state $|\epsilon_i \rangle$ to state $|\epsilon_k \rangle$
\begin{equation}\label{Gamma}
\Gamma^{n}_{i,k} \equiv \lambda\w_{ik}\left[ \left(1 + n_{n}(\w_{ik})\right) \rho_{ii} - n_{n}(\w_{ik}) \rho_{kk}\right] ~,
\end{equation}
defined for $i > k$;  for $i < k$, we have $\Gamma^{n}_{i,k} = - \Gamma^{n}_{k,i}$. Again, the index $n$ runs in $\{L, R\}$. We have
\begin{align}
\dot{\rho}_{1,1} &= 0 = \Gamma^{L}_{6,1} - \Gamma^{R}_{1,4}~,~~~~\dot{\rho}_{2,2} = 0 = -\Gamma^{L}_{2,5} + \Gamma^{R}_{3,2} ~,\nonumber\\
\dot{\rho}_{3,3} &= 0 = \Gamma^{L}_{7,3} - \Gamma^{R}_{3,2}~,~~~~\dot{\rho}_{4,4} = 0 = -\Gamma^{L}_{4,8} + \Gamma^{R}_{1,4} ~,\nonumber\\
\dot{\rho}_{5,5} &= 0 = \Gamma^{L}_{2,5} - \Gamma^{R}_{5,7}~,~~~~\dot{\rho}_{6,6} = 0 = -\Gamma^{L}_{6,1} + \Gamma^{R}_{8,6} ~,\nonumber\\
\dot{\rho}_{7,7} &= 0 = -\Gamma^{L}_{7,3} + \Gamma^{R}_{5,7}~,~~~~\dot{\rho}_{8,8} = 0 = \Gamma^{L}_{4,8} - \Gamma^{R}_{8,6} ~.
\end{align}
The system of equations above gives us 2 groups of 4 transitions, namely,
\begin{align}\label{grupo}
\Gamma^{L}_{6,1} &= \Gamma^{R}_{1,4} = \Gamma^{L}_{4,8} = \Gamma^{R}_{8,6} = \Gamma_{I},\nonumber\\
\Gamma^{L}_{2,5} &= \Gamma^{R}_{3,2} = \Gamma^{L}_{7,3} = \Gamma^{R}_{5,7} = \Gamma_{II} ~.
\end{align}

Finally, we compute the heat current. We have
\begin{align}
\cur_{L} &= \tr\{\Dop_{L}H_{S}\} = \sum_{j} \Dop_{j,j}^{L} H_{j,j}~,   \\
&= \Gamma^{L}_{6,1}H_{1,1} - \Gamma^{L}_{2,5}H_{2,2} + \Gamma^{L}_{7,3}H_{3,3} - \Gamma^{L}_{4,8}H_{4,4} \nonumber\\
&+ \Gamma^{L}_{2,5}H_{5,5} - \Gamma^{L}_{6,1}H_{6,6} - \Gamma^{L}_{7,3}H_{7,7} + \Gamma^{L}_{4,8}H_{8,8} \nonumber~,
\end{align}
where, to lighten the notation, we dropped out the index $S$ in $H$. It gives us
\begin{align}
\cur_{L} &= -\Gamma^{L}_{6,1}(H_{6,6} - H_{1,1}) + \Gamma^{L}_{2,5}(H_{5,5} - H_{2,2}) \nonumber\\
&- \Gamma^{L}_{7,3}(H_{7,7} - H_{3,3}) + \Gamma^{L}_{4,8}(H_{8,8} - H_{4,4}) \nonumber\\
&= \Gamma_{I}(-\w_{6,1} + \w_{8,4}) + \Gamma_{II}(\w_{5,2} - \w_{7,3}) = 0 ~,\nonumber
\end{align}
since $\w_{6,1} = \w_{8,4}$ and $\w_{5,2} = \w_{7,3}$.

There is no heat current in this system. It is interesting to remark that, many times, due to possible symmetries of the density matrix, the energy or the spin current in
several types of Heisenberg spin chains vanish, despite the existence of large boundary gradients \cite{PopLivi}.

We investigate other regimes and different cases of the system with $N=3$ and nearest-neighbor interactions, for example, the case with non-uniform external field and asymmetric interactions,
\begin{align}
H_{S} &= h\sop^{z}_{1} + (h+\zeta)\sop^{z}_{2} + (h+2\zeta)\sop^{z}_{3} \nonumber\\
&+ (\Delta + \delta)\sop^{z}_{1}\sop^{z}_{2} + (-\Delta + \delta)\sop^{z}_{2}\sop^{z}_{3}~.\label{H2}
\end{align}
In all these cases, the energy current is zero.

To follow, we increase the complexity of the system and add a next-nearest-neighbor interaction. We take
\begin{align}
H_{S} &= h\sop^{z}_{1} + (h+\zeta)\sop^{z}_{2} + (h+2\zeta)\sop^{z}_{3} \nonumber\\
&+ (\Delta + \delta)\sop^{z}_{1}\sop^{z}_{2} + (-\Delta + \delta)\sop^{z}_{2}\sop^{z}_{3} + \theta\sop^{z}_{1}\sop^{z}_{3}~.\label{H3}
\end{align}
[Details of the algebraic computations are presented in the Appendix.] Again, the transition rates are joined  into 2 groups of 4 terms,
$\Gamma_{I}$ and $\Gamma_{II}$ and a non-vanishing heat current appears, $ \cur_{L} = (\Gamma_{I} + \Gamma_{II})2^{2}\theta$. Making equal the transition rates,
$\Gamma_{I} = \Gamma_{II} \equiv \Gamma$, we obtain
\begin{align}
\cur = 2^{3}\Gamma \theta ~.
\end{align}

The value of $\Gamma$ is computed from Eqs. (\ref{grupo}) and (\ref{Gamma}), and from $\sum_{j}\rho_{j,j} = 1$. These expressions give us 9 linear equations involving 9 variables, namely,
$\Gamma$ and $\rho_{1,1}, \ldots, \rho_{8,8}$. Note that, from the definition of $n_{n}(\w)$, we can rewrite Eq.(\ref{Gamma}) shortly as
\begin{align}
\Gamma^{n}_{i,k}/\lambda = a^{n}_{k,i}\rho_{i,i} - a^{n}_{i,k}\rho_{k,k}~, ~~~~a^{n}_{k,i} \equiv \w_{k,i}n^{n}_{k,i}~.
\end{align}
Then, from Eq.(\ref{grupo}), we have the following 8 equations
\begin{align}
\Gamma/\lambda &= \Gamma^{L}_{6,1}/\lambda =  a^{L}_{1,6}\rho_{6,6} - a^{L}_{6,1}\rho_{1,1}~, \nonumber\\
\Gamma/\lambda &= \Gamma^{R}_{1,4}/\lambda =  a^{R}_{4,1}\rho_{1,1} - a^{R}_{1,4}\rho_{4,4}~, \nonumber\\
& \ldots \nonumber \\
\Gamma/\lambda &= \Gamma^{R}_{5,7}/\lambda =  a^{R}_{7,5}\rho_{5,5} - a^{R}_{5,7}\rho_{7,7}~,
\end{align}
In fact, we have 2 groups of 4 equations, which involve $\rho_{1,1}, \rho_{2,2}, \rho_{4,4}, \rho_{6,6}$ and $\rho_{5,5}, \rho_{3,3}, \rho_{8,8}, \rho_{7,7}$.
Details in the Appendix. We analyze the solution in the regime $\zeta > h , \Delta , \delta , \theta$. The occurrence of rectification is
clear: fixing $\beta_{L}$, i.e., $T_{L}$, and taking the limit of $\beta_{R} \rightarrow \infty$ ($T_{R}\rightarrow 0$) we obtain $\Gamma \rightarrow 0$. It means, no current from the left to the right side. However,
when we invert the baths, i.e., $\beta_{R}$ fixed and $\beta_{L} \rightarrow \infty$, for $h = \theta - \Delta - \delta$ [$\theta = h + \Delta + \delta$] we have a nonvanishing $\Gamma$. In other words, we can obtain a perfect diode (or a  perfect rectification) \cite{comentario}.
Otherwise, for different $h$,
$\Gamma$ also vanishes when $\beta_{L} \rightarrow \infty$, but much slower than in the opposite case as $\beta_{R} \rightarrow \infty$. Anyway, it means occurrence of thermal rectification for one side linked
to a bath in low temperature.

Now, we investigate the case $N=4$. Again, in any situation for the system with only nearest-neighbor interactions (asymmetry in the external field and/or in the interactions), we do not have heat current. Then, we turn to system
with next-nearest-neighbor interaction, that has a non-zero current and rectifies for $N=3$. Now, we take
\begin{align}
H_{S} &= h\sop^{z}_{1} + (h+\zeta)\sop^{z}_{2} + (h+2\zeta)\sop^{z}_{3} + (h+4\zeta)\sop^{z}_{3} \nonumber\\
&+ (\Delta + \delta)\sop^{z}_{1}\sop^{z}_{2} + (-\Delta + \delta)\sop^{z}_{2}\sop^{z}_{3} + (-3\Delta + \delta)\sop^{z}_{3}\sop^{z}_{4} \nonumber\\
&+ \theta \sop^{z}_{1}\sop^{z}_{3} + \phi\sop^{z}_{2}\sop^{z}_{4} ~.\label{H4}
\end{align}
And, again, after an easy but tedious algebra, we find that there is no heat current.

Following the strategy of introducing more intricate terms into the interaction until we find rectification,
we go beyond next-nearest-neighbor interaction and add, in the Hamiltonian above, the term
\begin{align}
\gamma\sop^{z}_{1}\sop^{z}_{4}~.
\end{align}
Then, after finding the energy eingenvectors and eigenvalues, and computing the possible transitions, which are joined in 4 groups of 4 elements due to the equations for the steady state $\dot{\rho}_{jj} = 0$, i.e., after
an easy but considerable algebra, we get
\begin{align}
\cur = 2^{4}\Gamma \gamma~.
\end{align}

Returning to the analysis of $N=4$, let us take, again, the regime of large $\zeta$, i.e., $\zeta > h, \Delta, \theta, \phi, \gamma$. It makes easy to identify the occurrence of thermal rectification. See Appendix for details. We have $\Gamma \rightarrow 0$ as $\beta_{R} \rightarrow \infty$ ($T_{R} \rightarrow 0$): no heat flow in the direction left to right. And, as in the scenario described for $N=3$, as $\beta_{L}
\rightarrow \infty$ there is the possibility of a perfect diode, for $h= \gamma -(\Delta + \delta + \theta)$ [i.e., $\gamma = h + \Delta + \delta + \theta$], which means a nonzero $\Gamma$. Otherwise, i.e., for other parameter
relations, $\Gamma$ goes to zero, but goes
slower than in the case of $\beta_{R} \rightarrow \infty$, showing anyway the occurrence of rectification.

In the Appendix, we argue to show that, for an asymmetric quantum Ising chain with $N$ sites and long range interaction, the heat current (when non-vanishing) is
\begin{align}
\cur = 2^{N}\Gamma\gamma_{N} ~,
\end{align}
where $\gamma_{N}$ is the interaction between $\sop^{z}_{1}$ and $\sop_{N}^{z}$. Moreover, following the algebraic formalism detailed carried out for small $N$, we predict the existence of at least a region of parameters with a perfect rectification, precisely, for
\begin{align}
\gamma_{N} = h + \gamma_{2} + \gamma_{3} + \ldots +\gamma_{N-1}~,
\end{align}
i.e., $h = \gamma_{N} - (\gamma_{2} + \gamma_{3} + \ldots +\gamma_{N-1})$, where, in our previous notation, $\gamma_{2} = \Delta + \delta$ and $\gamma_{3} = \theta$.

\section{Conclusions}

 We have to make some further remarks.

We believe that the necessity of an interaction linking the first spin in the chain to the last one comes to break some hidden symmetry preventing the appearance of the heat current. As already recalled, due to symmetries in the
Lindblad equation and density matrix, some Heisenberg spin chains are shown to have a zero spin or energy current even in the presence of strong boundary gradients \cite{PopLivi}.
That is, we believe that the ingredients for
rectifications are the quantum dynamics and the quantum nature of the baths, which bring temperature dependence to the bulk of the system, and the structural asymmetry, here present in the inter-site interactions. Even for the occurrence of a giant
heat rectification, no intricate interaction is necessary, in contrast with the usual classical model of oscillators.
It makes ubiquitous the occurrence of
thermal rectification in asymmetric quantum spin systems. Moreover, as shown, some spin models are capable to present a perfect rectification.

Concerning the experimental realizations of such spin chains, it is worth to recall that
 Heisenberg models can be implemented or simulated
  by means of cold atoms in optical lattices \cite{bloch2012quantum} or trapped
ions \cite{blatt2012quantum}. In Ref.\cite{trotzky2008}, for example, the model is
implemented in an optical double well, with a direct superexchange interaction between the spin of
the particles in different sites. In Ref.\cite{hauke2010complete}, a complicate (long range) Hamiltonian  is simulated for chains of up to 100 pseudo-spins. In Ref.\cite{duan2003controlling,whitlock2017simulating,PhysRevX.8.011032},
$XXZ$ models are involved in experiments with Rydberg atoms in optical traps.
Moreover, it is possible to engineer $XXZ$
 systems with different values for the $\sop^{x}_{j}\sop^{x}_{j+1}, \sop^{y}_{j}\sop^{y}_{j+1}$ and  $\sop^{z}_{j}\sop^{z}_{j+1}$ coefficients \cite{endres2016atom,barredo2016atom}.

To conclude, the appearance of heat rectification, much better, of a huge heat rectification, in these simple, experimentally realizable, quantum spin systems certainly sheds light on this important and difficult issue:
the building of efficient thermal diodes. We are confident that such a simplified itinerary will stimulate more research on the subject.

\section{Appendix}

In this section, we briefly describe some computation omitted from the main text. We believe that it may be useful to the reader to understand the involved algebra.

{\bf Solution for $N=3$}. In the case of $N=3$, the linear equations for the 2 groups  $\rho_{1,1}, \rho_{2,2}, \rho_{4,4}, \rho_{6,6}$ and  $\rho_{3,3}, \rho_{5,5}, \rho_{8,8}, \rho_{7,7}$, and $\Gamma$ are given by
\begin{align}
\alpha = &-a_{2,1}^{L}\rho_{1,1}& + &a_{1,2}^{L}\rho_{2,2}& + &0& + &0& \nonumber \\
\alpha = &+a_{4,1}^{R}\rho_{1,1}& + &0& - &a_{1,4}^{R}\rho_{4,4}& + &0& \nonumber \\
\alpha = &~0& + &0& + &a_{6,4}^{L}\rho_{4,4}& - &a_{4,6}^{L}\rho_{6,6}& \nonumber \\
\alpha = &~0& - &a_{6,2}^{R}\rho_{2,2}& + &0& + &a_{2,6}^{R}\rho_{6,6}&  ~,
\end{align}
\begin{align}
\alpha = &-a_{3,5}^{R}\rho_{5,5}& + &a_{5,3}^{R}\rho_{3,3}& + &0& + &0& \nonumber \\
\alpha = &+a_{8,5}^{L}\rho_{5,5}& + &0& - &a_{5,8}^{L}\rho_{8,8}& + &0& \nonumber \\
\alpha = &~0& + &0& + &a_{7,8}^{R}\rho_{8,8}& - &a_{8,7}^{R}\rho_{7,7}& \nonumber \\
\alpha = &~0& - &a_{7,3}^{L}\rho_{3,3}& + &0& + &a_{3,7}^{L}\rho_{7,7}&  ~,
\end{align}
where $\alpha \equiv \Gamma/\lambda$. Moreover, we have $\sum_{k=1}^{8}\rho_{k,k} = 1$. Note that the first system becomes exactly equal to the second one if we make the change of indices $1\leftrightarrow 5$, $2\leftrightarrow 3$, $4\leftrightarrow 8$, $6\leftrightarrow 7$ and $L \leftrightarrow R$.

The solution of $\rho_{k,k}$ in terms of $\Gamma$ is given by Cramer's formulas. For the determinant of the coefficients of first matrix we get
\begin{align}
{\rm detcoef_{1}} = &a^{L}_{2,1}a^{L}_{6,4}a^{R}_{6,2}a_{4,1}^{R} e^{\beta_{L}2(h-\Delta-\delta)} e^{\beta_{R}2(h+\Delta+2\zeta-\delta)}\nonumber\\
	&\times 2\sinh[2\theta(\beta_{L} - \beta_{R})]~,
\end{align}
where we have written $a_{k,j}$ above always with $k>j$. The frequencies involved in the expression are
\begin{align}
\w_{6,4} =& 2(h - \Delta - \delta + \theta)~,~~~~\w_{4,1} = 2(h + \Delta  + 2\zeta - \delta - \theta) ~, \nonumber \\
\w_{2,1} =& 2(h - \Delta - \delta - \theta)~,~~~~\w_{6,2} = 2(h + \Delta  + 2\zeta - \delta + \theta) ~.
\end{align}
Similarly, for the second matrix,
\begin{align}
{\rm detcoef_{2}} = &a^{L}_{7,3}a^{L}_{8,5}a^{R}_{5,3}a_{8,7}^{R} e^{\beta_{L}2(h+\Delta+\delta)} e^{\beta_{R}2(h-\Delta+2\zeta +\delta)}\nonumber\\
&\times 2\sinh[2\theta(\beta_{L} - \beta_{R})]~,
\end{align}
with the frequencies
\begin{align}
\w_{8,5} =& 2(h + \Delta + \delta + \theta)~,~~~~\w_{5,3} = 2(h - \Delta + 2\zeta + \delta - \theta) ~, \nonumber \\
\w_{7,3} =& 2(h + \Delta  +  \delta - \theta)~,~~~~\w_{8,7} = 2(h - \Delta  + 2\zeta + \delta + \theta) ~.
\end{align}

Solving the equations for $\rho_{k,k}$, we obtain
\begin{align}
&\rho_{1,1} + \rho_{2,2} + \rho_{4,4} + \rho_{6,6} = \frac{-\alpha}{\rm detcoef_{1}} R_{1} ~,\\
&R_{1} = \left\{ g_{1,2}^{L}a_{4,6}^{L}a_{1,4}^{R}  + a_{1,2}^{L}g_{6,4}^{L}a_{4,1}^{R} + g_{1,2}^{L}a_{6,4}^{L}a_{2,6}^{R} \right.\nonumber \\
&+ a_{2,1}^{L}g_{4,6}^{L}a_{6,2}^{R}
+ a_{1,2}^{L}g_{1,4}^{R}a_{2,6}^{R} +
a_{2,1}^{L}a_{1,2}^{R}g_{2,6}^{R} \nonumber \\
&\left. + a_{4,6}^{L}g_{1,4}^{R}a_{6,2}^{R} +  a_{6,4}^{L}g_{2,6}^{R}a_{4,1}^{R}\right\} \nonumber ~,
\end{align}
where
\begin{align}
g_{i,k}^{n} \equiv \w_{i,k}\coth\left(\frac{\beta_{n}\w_{i,k}}{2}\right) ~.
\end{align}
And similarly
\begin{align}
&\rho_{5,5} + \rho_{3,3} + \rho_{8,8} + \rho_{7,7} = \frac{-\alpha}{\rm detcoef_{2}} R_{2} ~,\\
&R_{2} = \left\{ g_{5,3}^{R}a_{8,7}^{R}a_{5,8}^{L}  + a_{5,3}^{R}g_{7,8}^{R}a_{8,5}^{L} + g_{5,3}^{R}a_{7,8}^{R}a_{3,7}^{L} \right.\nonumber \\
&+ a_{3,5}^{R}g_{8,7}^{R}a_{7,3}^{L}
+ a_{5,3}^{R}g_{5,8}^{L}a_{3,7}^{L} +
a_{3,5}^{R}a_{5,3}^{L}g_{3,7}^{L} \nonumber \\
&\left. + a_{8,7}^{R}g_{5,8}^{L}a_{7,3}^{L} +  a_{7,8}^{R}g_{3,7}^{L}a_{8,5}^{L}\right\} \nonumber ~.
\end{align}

From the expressions above and $\sum_{k=1}^{8}\rho_{k,k} = 1$, we obtain
\begin{align}
1 =& -\frac{\Gamma}{\lambda}\left\{ \frac{R_{1}}{\rm detcoef_{1}} + \frac{R_{2}}{\rm detcoef_{2}}\right\} ~,\nonumber \\
\Rightarrow \Gamma =& \frac{-\lambda}{\frac{R_{1}}{\rm detcoef_{1}} +
\frac{R_{2}}{\rm detcoef_{2}}}~.
\end{align}

To study the possibility of thermal rectification, we compare the limits
$\beta_{R} \rightarrow \infty$ (and $\beta_{L}$ finite), and the opposite situation $\beta_{L} \rightarrow \infty$ ($\beta_{R}$ finite). Considering the expressions and the frequencies involved, we have
\begin{align*}
{\rm detcoef_{1}} &\sim e^{4\zeta\beta_{R}}\frac{1}{e^{4\zeta\beta_{R}}}
\frac{1}{e^{4\zeta\beta_{R}}} \longrightarrow_{\beta_{R}\rightarrow \infty} 0 \\
{\rm detcoef_{2}} &\sim e^{4\zeta\beta_{R}}\frac{1}{e^{4\zeta\beta_{R}}}
\frac{1}{e^{4\zeta\beta_{R}}} \longrightarrow_{\beta_{R}\rightarrow \infty} 0
\end{align*}

It is easy to see that, as $\beta_{R} \rightarrow \infty$,  some terms in $R_{1}$ and $R_{2}$ stay finite, while
other ones go to zero. In short,
\begin{align*}
\Gamma \longrightarrow_{\beta_{R}\rightarrow \infty}  \frac{\lambda}{\infty} = 0 ~,
\end{align*}
that is, the heat current vanishes.

Now we turn to the analysis of inverted baths, i.e., $\beta_{L} \rightarrow \infty$ and $\beta_{R}$ finite.

We have
\begin{align*}
{\rm detcoef_{1}} &\sim \frac{1}{e^{2\beta_{L}(h-\Delta-\delta)}}\sinh(\beta_{L}2\theta) c_{1}~,\\
{\rm detcoef_{2}} &\sim \frac{1}{e^{2\beta_{L}(h+\Delta+\delta)}}\sinh(\beta_{L}2\theta) c_{2}^,
\end{align*}
where $c_{1}$ and $c_{2}$ do not depend on $\beta_{L}$. Again, some terms in $R_{1}$ and $R_{2}$ go to zero, but other ones
remain finite (non-zero). Hence, for $\theta = h + \Delta + \delta$, as $\beta_{L}\rightarrow \infty$ we have ${\rm detcoef_{1}}\rightarrow \infty$, and  ${\rm detcoef_{2}}\rightarrow C$, i.e., $\Gamma \rightarrow C'$. Where
$C$ and $C'$ are non-zero terms which do not depend on $\beta_{L}$.

It means, precisely, the occurrence of a perfect rectification. In fact, for other relations between $\theta, h, \Delta$ and $\delta$, if we stay in the regime of large $\zeta$, then $\Gamma$ may go to zero as $\beta_{L} \rightarrow \infty$, but it goes much slower than as $\beta_{R} \rightarrow \infty$. In other words, for one side of the chain at low temperature (non-zero), we still have huge thermal rectification.

{\bf Solution for $N=4$.} The formalism and manipulation follow the smaller case $N=3$. Now, the equations for the 16 $\rho_{k,k}$ are given in 4 groups of 4 equations, namely $k = 1, 2, 5, 9$; $k= 3, 6, 10, 13$;
$k= 4, 7, 11, 14$; and $k= 8, 12, 15, 16$. And the 16 equations involving $\Gamma^{n}_{j,k}$ are the following
\begin{align*}
\Gamma_{I} =& \Gamma^{L}_{2,1} = \Gamma^{R}_{1,5} = \Gamma^{R}_{9,2} = \Gamma^{L}_{5,9}~, \\
\Gamma_{II} =& \Gamma^{L}_{6,3} = \Gamma^{R}_{3,10} = \Gamma^{R}_{13,6} = \Gamma^{L}_{10,13}~, \\
\Gamma_{III} =& \Gamma^{L}_{7,4} = \Gamma^{R}_{4,11} = \Gamma^{L}_{11,14} = \Gamma^{R}_{14,7}~, \\
\Gamma_{IV} =& \Gamma^{L}_{12,8} = \Gamma^{R}_{8,15} = \Gamma^{R}_{16,12} = \Gamma^{L}_{15,16}~.
\end{align*}
We take $\Gamma_{I} = \Gamma_{II} = \Gamma_{III} = \Gamma_{IV} \equiv \Gamma$. Of course, we still have the equation $\sum_{k=1}^{16}\rho_{K,k} = 1$. Again, using the notation $\alpha \equiv \Gamma/\lambda$, for the first
group of equations, we have, for $\rho_{1,1}, \rho_{2,2}, \rho_{5,5}, \rho_{9,9}$,
\begin{align}
\alpha = &-a_{2,1}^{L}\rho_{1,1}& + &a_{1,2}^{L}\rho_{2,2}& + &0& + &0& \nonumber \\
\alpha = &+a_{5,1}^{R}\rho_{1,1}& + &0& - &a_{1,5}^{R}\rho_{5,5}& + &0& \nonumber \\
\alpha = &~0& + &0& + &a_{9,5}^{L}\rho_{5,5}& - &a_{5,9}^{L}\rho_{9,9}& \nonumber \\
\alpha = &~0& - &a_{9,2}^{R}\rho_{2,2}& + &0& + &a_{2,9}^{R}\rho_{9,9}&  ~,
\end{align}
that is equal to the first matrix for $N=3$, after the indices change $1,2,4,6 \leftrightarrow 1,2,5,9$. Let us denote the matrix of the coefficients above by $A_{1}$. The other sets of equations have similar expressions.
Precisely, the matrices are the same after the indices change
\begin{align*}
 A_{1} \longleftrightarrow A_{2} &\longleftrightarrow A_{3} \longleftrightarrow A_{4}  \\
 (1,2,5,9) \leftrightarrow (3,6,10,13) &\leftrightarrow (4,7,11,14) \leftrightarrow (8,12, 15, 16) ~.
\end{align*}

Performing the computation of the Cramer's formulas, we obtain
\begin{equation}
\Gamma = \frac{-\lambda}{\frac{X_{1}}{\rm detcoef}_{A_{1}} +
\frac{X_{2}}{\rm detcoef}_{A_{2}} + \frac{X_{3}}{\rm detcoef}_{A_{3}} + \frac{X_{4}}{\rm detcoef}_{A_{4}}}~.
\end{equation}
We do not make explicit the expressions for $X_{1}, \ldots, X_{4}$, which, similarly to the previous $R_{1}$ and $R_{2}$, remain finite as
$\beta_{R}$ or $\beta_{L} \rightarrow \infty$. The expressions for ${\rm detcoef}_{A_{k}}$ are described below.
\begin{align}
{\rm detcoef}_{A_{1}} =& a_{2,1}^{L}a^{L}_{9,5}a^{R}_{9,2}a^{R}_{5,1}e^{\beta_{L}2(h-\delta-\Delta-\theta)} \nonumber\\
 &\times e^{\beta_{R}2(h+ 4\zeta-\delta+3\Delta-\phi)}2\sinh[2\gamma(\beta_{L}-\beta_{R})] ~,\nonumber \\
 {\rm detcoef}_{A_{2}} =& a_{6,3}^{L}a^{L}_{13,10}a^{R}_{13,6}a^{R}_{10,3}e^{\beta_{L}2(h+\delta+\Delta-\theta)} \nonumber\\
 &\times e^{\beta_{R}2(h+ 4\zeta-\delta+3\Delta+\phi)}2\sinh[2\gamma(\beta_{L}-\beta_{R})] ~,\nonumber \\
 {\rm detcoef}_{A_{3}} =& a_{7,4}^{L}a^{L}_{14,11}a^{R}_{14,7}a^{R}_{11,4}e^{\beta_{L}2(h-\delta-\Delta+\theta)} \nonumber\\
 &\times e^{\beta_{R}2(h+ 4\zeta +\delta -3\Delta -\phi)}2\sinh[2\gamma(\beta_{L}-\beta_{R})] ~, \nonumber\\
 {\rm detcoef}_{A_{4}} =& a_{12,8}^{L}a^{L}_{16,15}a^{R}_{16,12}a^{R}_{15,8}e^{\beta_{L}2(h+\delta+\Delta+\theta)} \nonumber\\
 &\times e^{\beta_{R}2(h+ 4\zeta+\delta -3\Delta +\phi)}2\sinh[2\gamma(\beta_{L}-\beta_{R})] ~.
\end{align}
The involved frequencies are
\begin{align}
\w_{2,1} =& 2(h - \Delta - \delta - \theta -\gamma)~,\nonumber \\
\w_{5,1} =& 2(h + 4\zeta - \delta +3\Delta - \phi -\gamma) ~, \nonumber \\
\w_{9,5} =& 2(h - \Delta  - \delta - \theta + \gamma)~,\nonumber \\
\w_{9,2} =& 2(h + 4\zeta - \delta +3\Delta - \phi +\gamma) ~,\nonumber\\
\w_{13,10} =& 2(h + \Delta + \delta - \theta +\gamma)~,\nonumber \\
\w_{13,6} =& 2(h + 4\zeta - \delta +3\Delta + \phi +\gamma) ~, \nonumber \\
\w_{6,3} =& 2(h + \Delta  + \delta - \theta - \gamma)~,\nonumber \\
\w_{10,3} =& 2(h + 4\zeta - \delta +3\Delta + \phi -\gamma) ~,\nonumber\\
\w_{14,11} =& 2(h - \Delta - \delta + \theta +\gamma)~,\nonumber \\
\w_{11,4} =& 2(h + 4\zeta + \delta -3\Delta - \phi -\gamma) ~, \nonumber \\
\w_{7,4} =& 2(h - \Delta  - \delta + \theta - \gamma)~,\nonumber \\
\w_{14,7} =& 2(h + 4\zeta + \delta -3\Delta - \phi +\gamma) ~,\nonumber\\
\w_{16,15} =& 2(h + \Delta + \delta + \theta +\gamma)~,\nonumber \\
\w_{15,8} =& 2(h + 4\zeta + \delta -3\Delta + \phi -\gamma) ~, \nonumber \\
\w_{12,8} =& 2(h + \Delta  + \delta + \theta - \gamma)~,\nonumber \\
\w_{16,12} =& 2(h + 4\zeta + \delta -3\Delta + \phi +\gamma) ~.
\end{align}

Now, we examine the possibility of thermal rectification. First, we fix $\beta_{L}$ and take $\beta_{R}\rightarrow \infty$.
We have
\begin{align*}
{\rm detcoef}_{A_{k}} &\sim e^{8\zeta\beta_{R}}\frac{1}{e^{8\zeta\beta_{R}}}
\frac{1}{e^{8\zeta\beta_{R}}} \longrightarrow_{\beta_{R}\rightarrow \infty} 0 ~,
\end{align*}
where $k=1,2,3,4$. Consequently, $\Gamma \rightarrow 0$ as $\beta_{R}\rightarrow \infty$.
There is no heat current from the left to right side, as $T_{R} \rightarrow 0$.

Now we fix $\beta_{R}$. The asymptotic behavior as $\beta_{L}$ increases is
\begin{align}
{\rm detcoef}_{A_{1}} \sim \frac{1}{e^{\beta_{L}2(h-\Delta-\delta-\theta)}} e^{\beta_{L}2\gamma} ~,\nonumber\\
{\rm detcoef}_{A_{2}} \sim \frac{1}{e^{\beta_{L}2(h+\Delta+\delta-\theta)}} e^{\beta_{L}2\gamma} ~,\nonumber\\
{\rm detcoef}_{A_{3}} \sim \frac{1}{e^{\beta_{L}2(h-\Delta-\delta+\theta)}} e^{\beta_{L}2\gamma} ~,\nonumber\\
{\rm detcoef}_{A_{4}} \sim \frac{1}{e^{\beta_{L}2(h+\Delta+\delta+\theta)}} e^{\beta_{L}2\gamma} ~.
\end{align}
Hence, for $\gamma = h + \Delta + \delta + \theta$, as $\beta_{L}\rightarrow \infty$, we have that ${\rm detcoef}_{A_{4}}$ remains finite, and ${\rm detcoef}_{A_{1}}, {\rm detcoef}_{A_{2}}, {\rm detcoef}_{A_{3}} \rightarrow \infty$, and so, $\Gamma$ remains finite, which shows a perfect rectification. As previously argued in the case of $N=3$, for another relation between the involved parameters, we still have a huge rectification as one of the
sides is in very low temperature.

Following in details the derivation for the expressions for $N=2, 3, 4$ we may infer some formulas for generic $N$. For example, for the heat current,
\begin{align*}
\cur_{L} &= \tr\{\Dop_{L}H_{S}\} = \sum_{j} \Dop_{j,j}^{L} H_{j,j} \\
&= \sum_{i} \Gamma(-\w_{i_{1},i_{2}} + \w_{i_{3},i_{4}}) ~,
\end{align*}
where the first sum in $j$ involves $2^{N}$ terms, and the second one in $i$, $2^{N-2}$ terms. Each $\w$ in the sum, that is the difference between 2 terms $H_{j,j}$, involves a common term plus $\pm 2\gamma_{N}$, such that the difference between $\w$'s is $4\gamma_{N}$. Consequently,  $\cur_{L} = 2^{N}\Gamma\gamma_{N}$.

{\it Acknowledgments}:  E.P. was partially supported by CNPq (Brazil).



\begin{thebibliography}{100}
\bibitem{comentarios} We quote some authors. ``Rectification of heat transfer in nanodevices can be realized by combining the system inherant {anharmonicity} with structural asymmetry''\cite{SegalNitzan};
``the control of heat conduction by {nonlinearity} opens the possibility to propose new devices such as thermal rectifier'' \cite{Casati1}; ``our ability to play with {nonlinearities} allows us to think about these
possibilities [to build diodes]'' \cite{Casati2}; ``{anharmonic} interactions were, in particular pointed out as responsible for thermal rectification. While {it is obvious that harmonic systems do not rectify heat}, not all
anharmonic asymmetric systems do bring in the effect'' \cite{WuSegal}; ``the task of directing heat for information processing as in electronics requires a toolkit with suitable building blocks, namely, {nonlinear}
components that mimic the roles of diodes, \ldots'' \cite{LiRMP}.
\bibitem{comentario} It is worth mentioning that in the study of a junction, $N=2$,   we obtain $\cur = 2^{2}\Gamma\Delta$, where $\Delta$ is the interaction between the first and last spins (for $N=2$, it means the spin $1$ and
spin $2$).
The result seems to be slightly different from that found in
Ref.\cite{Werlang}, where the investigation of a junction is performed. The difference, however, disappears if we (as in \cite{Werlang}) introduce a factor $1/2$ in the definition of $H_{S}$, and a sign in $\Gamma$. The ocurrence of an optimal rectification is also described in
Ref.\cite{Werlang}, but there for $\beta_{L}\rightarrow\infty$, since their large magnetic field is in the left side, in opposition to our case here: our larger field is in the right side.
\end{thebibliography}
\end{document}